\documentclass[10pt,aps,prc,floatfix,twocolumn,nofootinbib, superscriptaddress,preprintnumbers]{revtex4-1}

\usepackage{amsfonts,amsmath,amssymb,bm, xspace}
\usepackage{color,graphicx}
\usepackage{bbm}
\usepackage{dcolumn}                 
\graphicspath{{./figures/}}

\usepackage{hyperref}                

\usepackage{array,physics}
\usepackage{dcolumn}
\usepackage[normalem]{ulem}
\newcolumntype{d}[1]{D{.}{.}{#1}}

\newcommand{\Msol}{\,\text{M}_{\odot}\xspace}

\newcommand{\fmiq}{\, \text{fm}^{-3}}

\newcommand{\MeV}{\, \text{MeV}}

\newcommand{\sat}{\textrm{sat}}
\newcommand{\nsat}{n_\sat}
\newcommand{\kink}{\textrm{kink}}
\newcommand{\tov}{\textrm{TOV}}

\hypersetup{%
    pdfsubject=Paper,
    pdfkeywords={nuclear physics} 
    unicode=true,
    breaklinks=true,
     colorlinks   = true,
     linkcolor = blue,
     citecolor = blue,
     menucolor = blue,
     urlcolor = blue
}

\begin{document}
\preprint{LA-UR-21-22340}

\title{Investigating Signatures of Phase Transitions in Neutron-Star Cores}

\author{R. Somasundaram}
\email{r.somasundaram@ipnl.in2p3.fr}
\affiliation{Univ Lyon, Univ Claude Bernard Lyon 1, CNRS/IN2P3, IP2I Lyon, UMR 5822, F-69622, Villeurbanne, France}

\author{I. Tews}
\affiliation{Theoretical Division, Los Alamos National Laboratory, Los Alamos, New Mexico 87545, USA}

\author{J. Margueron}
\affiliation{Univ Lyon, Univ Claude Bernard Lyon 1, CNRS/IN2P3, IP2I Lyon, UMR 5822, F-69622, Villeurbanne, France}

\date{\today}

\begin{abstract}
Neutron stars explore matter at the highest densities in the universe
such that their inner cores might undergo a phase transition from hadronic to exotic phases, e.g., quark matter.
Such a transition could be 
associated with non-trivial structures in the density behavior of the speed of sound such as jumps and sharp peaks.
Here, we employ a physics-agnostic approach to model the density dependence of the speed of sound in neutron stars and study to what extent the existence of non-trivial structures can be inferred from existing astrophysical observations of neutron stars.
For this, we exhaustively study different equations of state, including as well those with explicit first-order phase transitions.
We obtain a large number of different EoSs which reproduce the same astrophysical observations and obey the same physical constraints such as mechanical stability and causality. 
Among them, some have non-trivial structures in the sound speed while others do not.
We conclude that astrophysical information to date do \emph{not} require the existence of a phase transition to quark matter in the density range explored in the core of neutron stars.
\end{abstract}

\maketitle


\section{Introduction}

Recent multimessenger observations of Neutron stars (NSs), i.e., radio~\cite{Demorest:2010,Antoniadis:2013pzd,Cromartie:2019,Fonseca:2021wxt}, X-ray~\cite{Miller:2019,Riley:2019,Miller:2021qha,Riley:2021pdl}, gravitational-wave (GW) observations~\cite{TheLIGOScientific:2017,Abbott:2018exr}, and their electromagnetic (EM) counterparts~\cite{LIGOScientific:2017a,LIGOScientific:2017b}, have provided valuable new insights into the equation of state (EoS) of dense matter~\cite{Bauswein:2017vtn, Annala:2017llu, Most:2018hfd, Radice:2018ozg, Capano:2019eae,  Dietrich:2020lps, Legred:2021hdx,Miller:2021qha,Raaijmakers:2021uju}. 
Nevertheless, the composition of matter at several times nuclear saturation density ($\nsat \approx 0.16 \fmiq$) remains largely unknown~\cite{Lattimer:2004} but might be elucidated by future multimessenger data.

While NS matter at densities $n \approx \nsat$ is composed primarily of nucleons, a change in the degrees of freedom to exotic forms of matter, such as quark matter, might occur at larger densities inside NSs~\cite{Zdunik:2013,Alford:2013,Chamel:2013}.
Such a change is expected to likely manifest itself in terms of non-trivial structures in the sound-speed profile~\cite{Tan:2020,Tan:2021ahl}. 
For example, an abrupt First-Order Phase Transition (FOPT) creates a discontinuous drop in the sound speed as a function of density~\cite{Christian:2019qer,Christian:2018jyd}.
Alternatively, a softening of the EoS indicated by the sound speed approaching the conformal limit, $c_s\to 1/\sqrt{3}$, might be indicative of the onset of weakly coupled quark matter~\cite{Gorda:2018gpy,Gorda:2021znl}. 
In stark contrast to such softening phase transitions, a transition to quarkyonic matter could stiffen the EoS, leading to a sharp peak in the sound speed profile~\cite{McLerran:2018, Jeong:2019,Sen:2020,Duarte:2020, Zhao:2020, Sen:2020qcd,Margueron_prep}. 

Recently, studies investigated non-trivial structures in the EoS above saturation density, such as bumps in the sound speed~\cite{Tan:2020,Tan:2021ahl} or a kink in the EoS~\cite{Annala:2019}.
The authors of Ref.~\cite{Annala:2019}, using a general extension scheme in the speed of sound along with several theoretical and experimental constraints, claimed to have found evidence for a phase transition to quark matter in the heaviest NSs due to the presence of a kink in the envelope of all EoS models.
On the other hand, the authors of Ref.~\cite{Tan:2020} explicitly considered non-trivial structures in the speed of sound such as kinks, dips and peaks.
This allowed the authors to construct NSs the masses of which are consistent even with the mass of the secondary component of GW190814~\cite{LIGOScientific:2020zkf}. 
They concluded that such non-trivial structures are likely present at densities probed in very massive NSs ($\approx 2.5 \Msol$).
Note, however, that the secondary component in GW190814 was likely not a neutron star~\cite{Essick:2020ghc,Tews:2020ylw}.

Here, we re-investigate if non-trivial structures in the speed of sound can be inferred from present astrophysical data. 
For concreteness, we define a non-trivial structure to be either a non-monotonous dependence of the sound speed on the energy density or a tendency of the sound speed towards the conformal limit $1/\sqrt{3}$ in maximally massive stars. 
We use a systematic approach to model the EoS in the speed of sound vs. density plane, employing a piece-wise linear model for the speed of sound which is a modified version of the scheme of Ref.~\cite{Tews:2018iwm} and similar to Ref.~\cite{Annala:2019} but with a larger number of model parameters, see Sec.~II for details.
We then group different EoS realizations according to the slope in the speed of sound, the appearance of non-trivial structures, or explicit FOPT, and analyze the effect of astrophysical NS observations in Sec~III.
We compare our results with those presented in the literature and comment on the evidence for phase transitions linked to astrophysical data.
Our conclusions are given in Sec.~IV.

\section{Equation of State Model}

\begin{figure*}
    \centering
    \includegraphics[width=\textwidth]{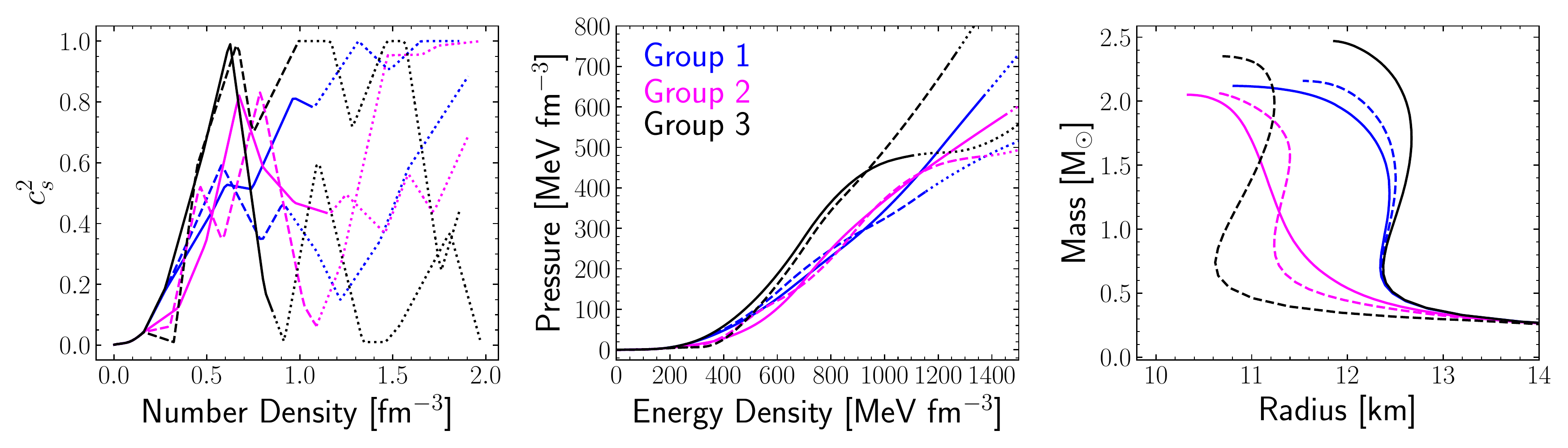}
    \caption{Schematic representation of the employed EoS model showing two EoSs per group.
    Different colors indicate the EoS group, and solid and dashed lines indicate the two EoSs in each group. 
    Dotted lines depict the EoS beyond the central density in the maximum-mass configuration, i.e., the unstable NS branch.}
    \label{fig:EOS model}
\end{figure*}

At low densities, up to nuclear saturation density $\nsat$, we fix our EoS to be given by the SLy4 energy-density functional~\cite{Chabanat:1997}, a phenomenological force that is well calibrated to nuclear matter as well as finite nuclei properties, and commonly used in astrophysical applications. 
Beyond $\nsat$, for each EoS we create a non-uniform grid in density between $\nsat$ and $12\nsat$ by randomizing an initial uniform grid with a spacing of $\nsat$: at each grid point, a density shift drawn from a uniform distribution between $-0.4\nsat$ and $0.4\nsat$ is added and defines the set $\{n_i\}_{i=1,11}$.
We then sample random values for $c_s^2(n_i)$ between $0$ and $c^2$ with $c$ being the speed of light (we set $c=1$ in the following).
Finally, we connect all points $c_{s,i}^2(n_i)$ using linear segments. The left panel in Fig.~\ref{fig:EOS model} illustrates our construction with a few examples. 

We then sort the resulting EoSs into three groups according to the maximal slope in the speed of sound $c'_{\rm max}={\rm max} \lbrace | dc^2_s/dn |\rbrace$. Defining the slope at $\nsat$ to be $c'_{\sat}= dc^2_s/dn (\nsat)=0.55~\text{fm}^{3}$,
\begin{itemize}
    \item group 1 contains EoSs whose maximal slope is less than three times the slope at $\nsat$, $c'_{\rm max}\leq 3c'_{\sat}$, 
    \item group 2 contains all EoSs with $3c'_{\sat} < c'_{\rm max}\leq 6c'_{\sat}$, 
    \item and group 3 contains all EoSs with $6c'_{\sat}< c'_{\rm max}\leq 9c'_{\sat}$\,. 
\end{itemize}
These groups are, thus, mutually exclusive.
The upper limit in the maximal speed of sound for group 3 allows us to disregard EoSs for which the sound speed strongly oscillates with density, a case which is not observed in any EoS models besides those which incorporate a FOPT.
We explicitly construct FOPTs later in the manuscript and analyze their impact on our results.
Note that $c_s^2$ is piece-wise linear in density, see left panel in Fig.~\ref{fig:EOS model}, which implies that $c'$ is piece-wise constant. 
However, our approach still allows for sufficient model freedom as we use 11 segments and the maximum $c'$ over all segments is used to partition the EoSs.
We have generated 10,000 EoSs in each group and a sample of two EoSs per group is shown in Fig.~\ref{fig:EOS model} to provide a schematic representation of our EoS model.

It is interesting to analyze the properties of the few example EoSs shown in Fig.~\ref{fig:EOS model}. 
Group 3 (back lines) collects some of the stiffest EoSs, allowing for large slopes of the sound speed, and therefore, a fast increase of pressure.
However, one of the EoSs (dashed black line) undergoes a softening at $2n_\sat$ just before stiffening again at larger densities. 
This low-density softening is particularly visible in the middle panel which shows the pressure as function of the energy density.
Finally, the last panel shows the corresponding mass-radius relations, where this particular  EoS predicts the smallest radii for canonical mass NSs. 
It, however, predicts larger radii for a $2 \Msol$ NS as compared to the examples from other groups. 
This is a consequence of the stiffening of the EoSs at densities larger than about $3n_\sat$.
Note that the other EoS from Group 3 predicts the largest radii for all masses due to the absence of such low-density softening.
Hence, the differentiation of the EoS in groups according to the slope of the speed of sound can not directly be mapped to NS radii, which is also obvious from the intersection of EoS belonging to different groups in the mass-radius plane.
Instead, this differentiation is non-trivial because all NS structure properties are integrated quantities.

In addition, the two EoSs of Group 3 exhibit a pronounced peak in the sound speed, centered around $\approx 4 \nsat$.
Investigating such non-trivial structures are the main goal of this work. 
The EoSs in Groups 1 and 2, on the other hand, have a more continuous behavior, with less drastic bumps and kinks.
Note, that individual EoS in each group can exhibit both trivial or non-trivial behavior.
Therefore, in the following, we will average over EoSs in each group to assess their behavior.

\begin{figure*}
    \centering
    \includegraphics[trim= 1.5cm 0.1cm 0.1cm 0.3cm, clip=,width=\textwidth]{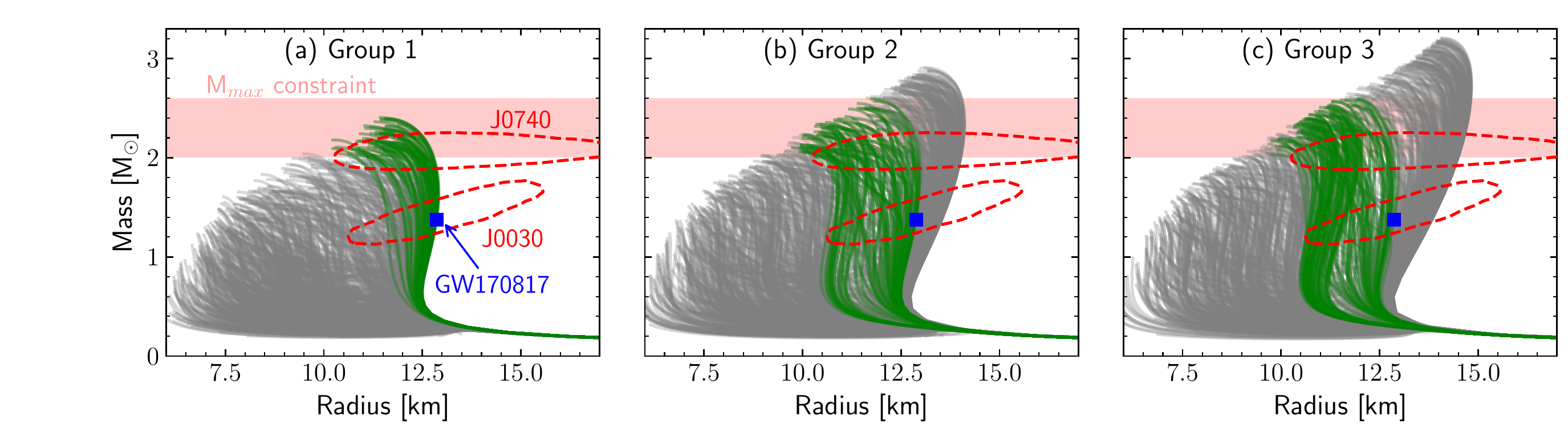}
    \caption{Mass-radius curves for all EoSs that we employ in our work (gray). The samples are divided into three panels corresponding to three EoS groups (see text). 
    Note that for each group only 1000 random samples out of the total 10000 EoSs are shown.
    We also show the observational constraints enforced in this work.
    EoSs that pass all observational constraints at the 90\% confidence level are shown in green. See text for more details.}
    \label{fig:prior}
\end{figure*}

The density dependent sound speed can be used to obtain the EoS. This is done by inverting the expression $c_s^2=dp/d\epsilon=(n d\mu)/(\mu dn)$ to obtain the chemical potential $\mu$, pressure $p$, and energy density $\epsilon$ in the interval $n_i\le n\le n_{i+1}$,
\begin{eqnarray}
\log (\frac{\mu(n)}{\mu_i} ) &=& \int_{n_i}^n \frac{c^2_s(n')}{n'} dn'\, , \\
p(n) &=& p(n_i) + \int_{n_i}^{n} c^2_s(n') \mu(n') dn' \,, \\
\epsilon(n) &=& \epsilon(n_i) + \int_{n_i}^{n} \mu(n') dn' \, .
\end{eqnarray}
Finally, we solve the TOV equations for each EoS to determine NS radii ($R$) and dimensionless tidal deformabilities ($\Lambda$) as functions of masses ($M$), see for instance Ref.~\cite{Tews:2018iwm} for more details.

In Fig.~\ref{fig:prior}, we show the resulting mass-radius curves for all three EoS groups together with the astrophysical observations that we consider in this work: 
\begin{itemize}
\item the NICER observations of millisecond pulsars J0030+0451~\cite{Miller:2019,Riley:2019} and J0740+6620~\cite{Miller:2021qha,Riley:2021pdl},
\item the LIGO gravitational-wave observation GW170817~\cite{TheLIGOScientific:2017,De:2018},
\item and upper and lower limits on the maximum NS mass, M$_{\tov}$\,.
\end{itemize}
For NICER's observation of J0740+6620, we have averaged over the analyses of Ref.~\cite{Miller:2021qha} and Ref.~\cite{Riley:2021pdl} and use the contours representing the 90\% CL. For the pulsar J0030+0451, we use the analysis of Ref.~\cite{Miller:2019} only. We do not average over the results of Ref.~\cite{Miller:2019} and Ref.~\cite{Riley:2019} since the two analyses are quite similar.
Then, an EoS is accepted if it passes through the contours, and it is rejected otherwise.

For GW170817, we have transformed the inferred result for the tidal deformability, $\tilde{\Lambda}=222^{+420}_{-138}$ at 90\% confidence level (CL)~\cite{De:2018}, into a single constraint on the radius for the mass $m=1.38 \Msol$.
We have calculated an upper bound on the radius at that mass by running over all our EoSs compatible with GW170817 and by fixing the mass ratio $q=m_1/m_2=1$. 
This upper bound on the radius at $1.38 \Msol$ is $12.9$~km and it is indicated by a blue dot in the figure. 

Finally, for M$_{\tov}$ we impose the constraint $2 \Msol < $M$_{\tov} < 2.6 \Msol$. 
The lower bound of $2\Msol$ is chosen to account for heavy pulsar radio observations~\cite{Demorest:2010, Antoniadis:2013pzd, Cromartie:2019, Fonseca:2021wxt}.
The upper bound of $2.6\Msol$ is consistent with the mass of the secondary object in the GW190814 event, which is likely a black hole~\cite{LIGOScientific:2020zkf,Tews:2020}.

We note that all observations are implemented by hard cuts at the 90\% CL, including the NICER results.
This approach is sufficient as we search for general trends of the speed of sound. 
However, detailed inferences of the EoS should use full posteriors for all astrophysical data.
The EoSs that satisfy all constraints at the 90\% CL are shown in green in Fig.~\ref{fig:prior}.

We note that additional NS radii have been inferred from X-ray observations~\cite{Al-Mamun:2020vzu}.
These include measurements of quiescent low-mass X-ray binaries~\cite{Rybicki:2005id} and NSs which exhibit photospheric radius expansion X-ray bursts~\cite{Ozel:2008kb}, see Refs.~\cite{Lattimer:2012nd,Ozel:2016oaf} for reviews. 
However, these measurements contain sizeable systematic uncertainties which prohibit a meaningful inclusion of these data in our analysis. 
We therefore restrict ourselves only to the previously mentioned data.

\section{Results}

We present the results of our analysis based on 10,000 EoSs per group, which are constrained by astrophysical data and physical constraints of causality and mechanical stability. Our present uncertainty on the density dependence of the EoS is explored.

\subsection{Discussion of the EoS}

\begin{figure}
    \centering
    \includegraphics[trim= 0 0.6cm 0 0, clip=,width=0.49\textwidth]{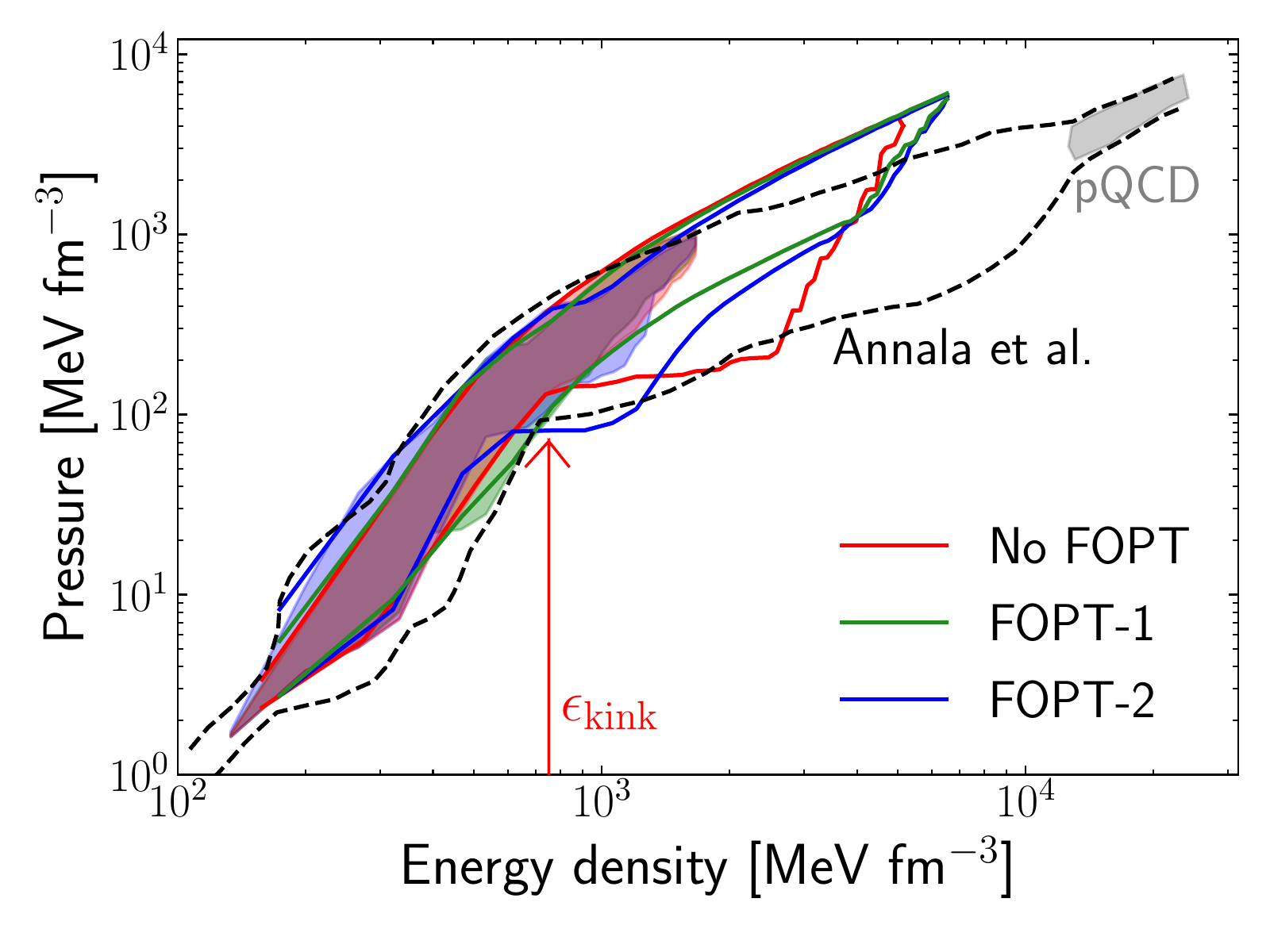}
    \caption{EoSs of this work that satisfy observational constraints. 
    We show envelopes for EoSs without FOPT (red) and EoSs with FOPT with different onset density ranges (green and blue), see details in the text. 
    The shaded bands correspond to stable NS configurations, whereas the solid lines show the EoSs extended beyond the maximally massive NS configurations.
    The black contour depicts the results of Ref.~\cite{Annala:2019}, and the gray contour represents the perturbative QCD constraint.}
    \label{fig:p_eps}
\end{figure}

The envelopes containing all EoSs that survive astrophysical constraints are shown in Fig.~\ref{fig:p_eps}.
The filled contours enclose EoSs on the stable NS branch, for which the density is limited by $n_\tov$, the central density of the maximum-mass NS with M=M$_\tov$, while the dashed contours enclose the full EoS up to the maximum density we considered ($12 n_\sat$). So the dashed contours also contain the extensions of the EoSs to the unstable branch. Fig.~\ref{fig:p_eps} illustrates that when we show EoSs at densities above $n_\tov$, we also find a kink similar to the one discussed in  Annala \textsl{et al.}~\cite{Annala:2019} at $\epsilon_\kink \gtrapprox 700~\MeV \fmiq$. 
More precisely, we observe that some EoSs can be arbitrarily soft above $\epsilon_\kink$, especially if this regime cannot be observed for NSs at equilibrium.
For these EoSs, the softening could be enough to destabilize the star, see also the discussion in Ref.~\cite{Annala:2019}. 
At even larger densities, our EoS envelopes broaden compared to Ref.~\cite{Annala:2019}, and include both softer and stiffer EoS.
Note, that unlike Ref.~\cite{Annala:2019}, we do not incorporate constraints from pQCD, which explains why stiffer EoS are allowed in the unstable branch in our case.
However, these differences have no impact on stable NSs, which remain inside the contour obtained in Ref.~\cite{Annala:2019}.
We also note that the constraining power of pQCD calculations has been found to be model-dependent: We have shown in a parallel work~\cite{Somasundaram:2022ztm} that pQCD calculations with their theoretical uncertainties, when imposed as suggested in Ref.~\cite{Komoltsev:2021}, do not further constrain the EoS once astrophysical data is taken into account but see Ref.~\cite{Gorda:2022jvk} for a different conclusion.

In Fig.~\ref{fig:p_eps}, we also show results for EoSs with an explicit FOPT, built upon each original EoS. 
We describe FOPT in terms of three parameters:
the transition energy density $\epsilon_t$, the width of the transition $\Delta \epsilon$, and a constant sound speed\footnote{While a \textit{constant} sound speed is not required, we do not believe that this approximation biases our results. The effects of density dependent sound speed profiles beyond the FOPT will be studied in a future work.} in the high-density phase beyond the phase transition, $c^2_{s,t}$~\cite{Zdunik:2013,Alford:2013,Chamel:2013}. 
For each EoS, random values are drawn from the uniform ranges $\epsilon_t = [400,800] \MeV \fmiq$, $\Delta \epsilon = [0.2,0.6] \epsilon_t$ and $c^2_{s,t} = [1/3,1]$, and the EoSs are tested against astrophysical constraints as before. 
We further separate the EoSs into two subgroups: $\epsilon_t = [400,600] \MeV \fmiq$ (FOPT-1) and $\epsilon_t = [600,800] \MeV \fmiq$ (FOPT-2).
For the group with larger onset density, we again observe a kink in the EoS envelope, while the other EoS group shows a smooth behavior.
The softening of the EoSs without FOPT observed in Fig.~\ref{fig:p_eps} at $\epsilon_\kink$
is somewhat similar to the softening of the FOPT EoSs with $\epsilon_t = [600,800] \MeV \fmiq$.
It is, therefore, tempting to conclude that this softening is a signal of a phase transition to exotic matter. 
In the following, we will investigate this point in detail by studying the behaviour of the adiabatic index $\gamma \equiv \frac{d\log p}{d\log \epsilon}= (\epsilon/p) \, c_s^2$ and the sound speed $c^2_s$ for all EoSs shown in Fig.~\ref{fig:p_eps}.

\subsection{Discussion of the existence of phase transitions as not trivial structure in the sound speed}

\begin{figure*}
\centering
\includegraphics[trim= 1.5cm 0.1cm 2.5cm 0.3cm, clip=,width=\textwidth]{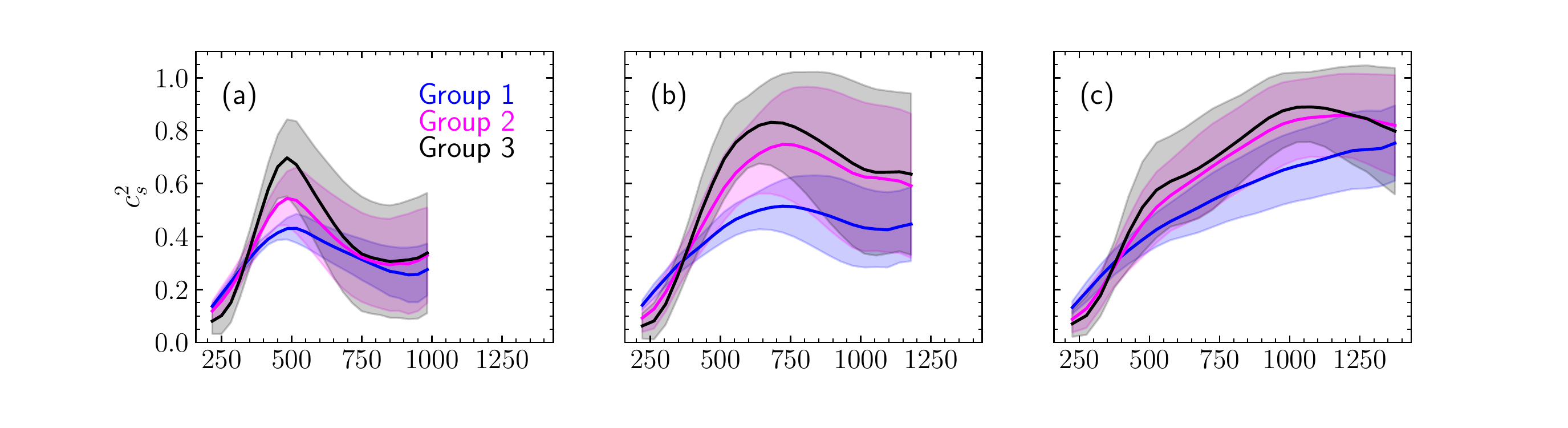}
\includegraphics[trim= 1.5cm 0.1cm 2.5cm 0.3cm, clip=,width=\textwidth]{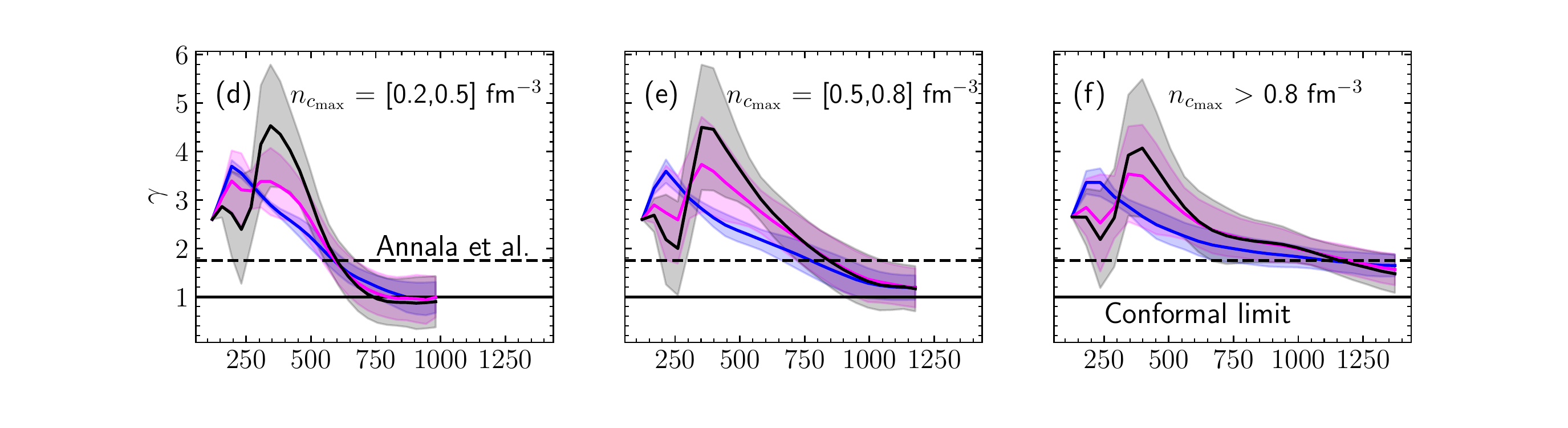}
\includegraphics[trim= 1.5cm 0.1cm 2.5cm 0.3cm, clip=,width=\textwidth]{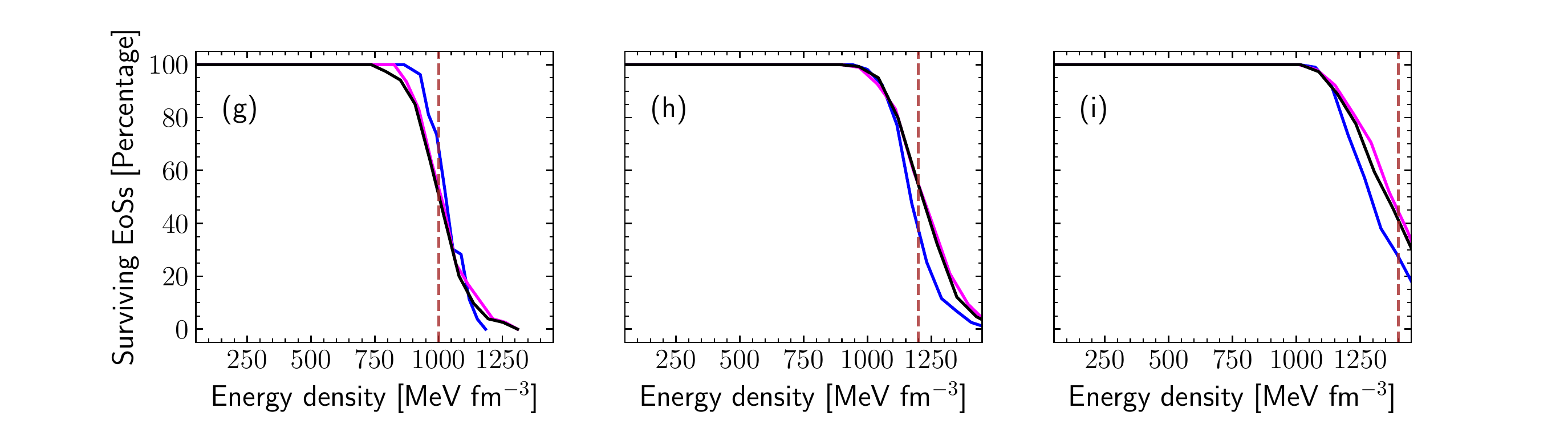}
\caption{The squared sound speed (top) and adiabatic index (middle) versus the energy density. We show the model-averaged bands for EoS that fulfill the astrophysical constraints (mean values with solid lines, band width representing the standard deviation).
The different panels correspond to different locations of the maximum of the sound speed. 
The fraction of surviving EoSs used for the average is shown in the bottom row and the energy densities where the bands are terminated are indicated with vertical brown lines (see text).}
\label{fig:posterior_bands}
\end{figure*}

In Fig.~\ref{fig:posterior_bands}, we show sound speed distributions as functions of energy density, for all EoSs fulfilling the observational constraints previously described. Regarding the distributions, the solid lines correspond to the sample mean of our EoS ensemble whereas the widths of the bands indicate the $2 \sigma$ uncertainty, where $\sigma$ is the sample standard deviation. The sample means and the standard deviations are computed at different points along the x-axis (after a suitable binning) to give the bands. 
Such an average over all EoSs in each group allows us to study general EoS trends and look for non-trivial structures.
Note that our analysis is restricted to the stable NS branch, implying that our EoSs are terminated at $n_\tov$, which corresponds to a different energy density for each EoS. 
This decreases the number of EoSs in the band as the energy density increases. The fraction of EoSs that survive as a function of the energy density is shown in the bottom row of Fig.~\ref{fig:posterior_bands}. The energy densities at which the bands are terminated are indicated by vertical brown lines in the bottom panels. 

We separate each EoS group into subgroups, shown in different panels, according to the density where the speed of sound reaches its maximum $n_{c_{\textrm{max}}}$.
Note that the energy density above which the number of EoSs starts decreasing depends primarily on the subgroup.
This value is around $\approx 800 \MeV \fmiq$ in panel (g), around $\approx 1000 \MeV \fmiq$ in panel (h) and around $\approx 1100 \MeV \fmiq$ in panel (i).  

As one can see in Fig.~\ref{fig:posterior_bands}, while some subgroups contain EoSs with clear peaks, astrophysical observations do \emph{not} require the EoS to have significant structures in the speed of sound, in particular a decrease to $c_s^2\approx 1/3$, see Fig.~\ref{fig:posterior_bands}(c).
The more pronounced peaks in Fig.~\ref{fig:posterior_bands}(a) are reminiscent of the quarkyonic model~\cite{McLerran:2018, Jeong:2019,Sen:2020,Duarte:2020, Zhao:2020, Sen:2020qcd,Margueron_prep} whereas the broader peaks in Fig.~\ref{fig:posterior_bands}(b) can potentially be interpreted as weaker change of phases. 
We have found that the peaks observed in Fig.~\ref{fig:posterior_bands} are well correlated with the upper limit on M$_{\tov}$: by increasing this upper limit, the peak is less pronounced and the sound speed $c_s^2(n)$ can remain high for large $n$. This is well explained by the fact that a peak in the sound speed implies a softening of the EoS for densities above the peak density.

\begin{figure*}[t]
    \centering
    \includegraphics[trim= 1.5cm 0.1cm 2.5cm 0.3cm, clip=,width=\textwidth]{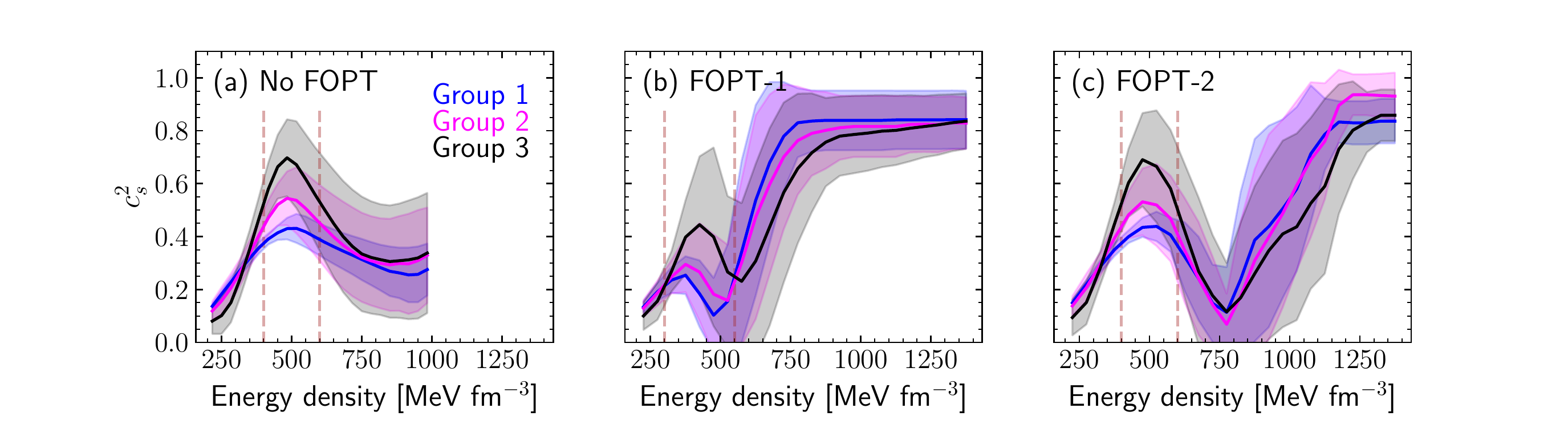}
    \caption{We compare the speed-of-sound bands for the various groups without a FOPT (left panel, similar to left panel of Fig.~\ref{fig:posterior_bands}) and when a FOPT is explicitly included. We show results for two ranges of the onset density of the FOPT. The dashed vertical lines encapsulate the region inside which a peak in the sound-speed occurs.}
    \label{fig:posterior_bands_fuse}
\end{figure*}

Our findings reported in Fig.~\ref{fig:posterior_bands} are consistent with those presented in Ref.~\cite{Tan:2021ahl}, which concluded that existing astrophysical data does not necessarily enforce a bump in $c^2_s$.
On the other hand, our results are in tension with the findings of Ref.~\cite{Annala:2019} which chose a value of the adiabatic index $\gamma=1.75$ to distinguish between hadronic and quark-matter models.
In the bottom panels of Fig.~\ref{fig:posterior_bands}, we find that basically all $\gamma$ bands drop below $\gamma=1.75$ at higher densities, for groups 1-3.
A straightforward comparison of the top and bottom panels in Fig.~\ref{fig:posterior_bands} however shows that while $\gamma$ asymptotically approaches $1$ in all cases, the sound speeds exhibit no preferred asymptotic value that can be identified as the conformal quark matter limit. 
We therefore stress that the reduction of the adiabatic index below $\gamma\approx 1.75$ is only a \emph{necessary but not a sufficient condition} for the appearance of quark matter.

We conclude this first part of the analysis that our careful modeling of the density dependence of the sound speed has not provided any evidence for quark-matter cores, as was claimed in  Ref.~\cite{Annala:2019}.
While we do not expect the inclusion of the pQCD EoS in our analysis to change this conclusion, further work is warranted.  

\subsection{Discussion of the existence of explicit first order phase transitions}

We finally analyze the impact of explicitly including a FOPT on the sound speed profiles. 
The results are shown in Fig.~\ref{fig:posterior_bands_fuse}, where we compare (panel (a)) EoSs without FOPT but predicting a significant peak (as shown in Fig.~\ref{fig:posterior_bands}(a)) with EoSs undergoing a FOPT. 
The EoSs with FOPTs are separated into two groups, FOPT-1 (panel (b)) and FOPT-2 (panel (c)). The two FOPT cases are distinguished by the energy density at which the phase transition occurs: $\epsilon_t = [400,600] \MeV \fmiq$ for FOPT-1 and $\epsilon_t = [600,800] \MeV \fmiq$ for FOPT-2.

When including a FOPT, we always observe the formation of a clear peak in the speed of sound profile, located at an energy density inside the region indicated by the dashed brown lines, similar to what we observe in Fig.~\ref{fig:posterior_bands_fuse}(a). 
The region is $[300:550]$ ($[400:600]$) MeV~fm$^{-3}$ for FOPT-1 (FOPT-2).
This suggests that if a FOPT occurs in dense matter in the parameter ranges explored in FOPT-1 and FOPT-2, it is preceded by a sharp increase in the sound speed above the conformal limit, reaching its maximum at approximately $400$-$500 \MeV \fmiq$. 
However, we stress once again that present astrophysical data does \emph{not} necessarily imply that the EoS undergoes a FOPT.

\section{Conclusions}

We have used a framework based on a speed-of-sound extension to address the question of phase transitions in dense matter in a systematic way. 
We have classified EoSs agreeing with present astrophysical data based on the behaviour of their sound speed profiles.
We were able to identify EoSs with peaks in the sound-speed profiles, resembling the behavior of quarkyonic matter, while for others we have explicitly included a FOPT.
However, our analysis also revealed the presence of (sub)groups of EoSs that have neither a clear peak nor an asymptotic tendency to the conformal limit. 
This leads us to conclude that present astrophysical data does \emph{not} favour exotic matter over the standard hadronic matter or a crossover feature. 
Nevertheless, our analysis also shows that, in the near future, new astrophysical data has the clear potential to be decisive, one way or the other, regarding the long-standing question of phase transitions in dense matter. 
We believe that the tools developed in this work -- by classifying the type of sound-speed density dependence in terms of simple properties -- paves the way for further research in this direction.

\acknowledgements

We thank V. Dexheimer, T. Dore, J. Noronha-Hostler,  J. Schaffner-Bielich, H. Tan, and N. Yunes for helpful feedback on the manuscript.
R.S. is supported by the PHAST doctoral school (ED52) of \textsl{Universit\'e de Lyon}. R.S. and J.M. are both supported by CNRS grant PICS-08294 VIPER (Nuclear Physics for Violent Phenomena in the Universe), the CNRS IEA-303083 BEOS project, the CNRS/IN2P3 NewMAC project, and benefit from PHAROS COST Action MP16214 and the PHAROS COST Action MP16214 as well as the LABEX Lyon Institute of Origins (ANR-10-LABX-0066) of the \textsl{Universit\'e de Lyon} for its financial support within the program \textsl{Investissements d'Avenir} (ANR-11-IDEX-0007) of the French government operated by the National Research Agency (ANR).
The work of I.T. was supported by the U.S. Department of Energy, Office of Science, Office of Nuclear Physics, under contract No.~DE-AC52-06NA25396, by the Laboratory Directed Research and Development program of Los Alamos National Laboratory under project numbers 20190617PRD1, 20190021DR, and 20220658ER, and by the U.S. Department of Energy, Office of Science, Office of Advanced Scientific Computing Research, Scientific Discovery through Advanced Computing (SciDAC) NUCLEI program.

\bibliography{biblio}
\end{document}